\documentclass[%
 reprint,
superscriptaddress,
 amsmath,amssymb,
 aps,prd
]{revtex4-2}

\usepackage{graphicx}
\usepackage{dcolumn}
\usepackage{bm}
	\newcommand{\ket}[1]{\left| #1 \right\rangle}
	\newcommand{\bra}[1]{\left\langle #1 \right|}
	
	\newcommand{\braket}[2]{\left\langle #1 \right. \left| #2 \right\rangle}

\begin{document}


\title{Hidden conformal symmetry on the black hole photon sphere
} 

\author{Bernard Raffaelli}
\email{bernard.raffaelli@u-bourgogne.fr}
\affiliation{Institut de Math\'{e}matiques de Bourgogne, UMR 5584 CNRS, Universit\'{e} Bourgogne Franche-Comt\'{e}, F-2100 Dijon, France}

\date{\today}

\begin{abstract}
We consider a class of static and spherically symmetric black hole geometries endowed with a photon sphere. On the one hand, we show that close to the photon sphere, a massless scalar field theory exhibits a simple dynamical $SL(2,\mathbb{R})$ algebraic structure which allows to recover the discrete spectrum of the weakly damped quasinormal frequencies in the eikonal approximation, and the associated quasinormal modes from the algebra representations. On the other hand, we consider the non-radial motion of a free-falling test particle, in the equatorial plane, from spatial infinity to the black hole. In the ultrarelativistic limit, we show that the photon sphere acts as an effective Rindler horizon for the geodesic motion of the test particle in the $(t,r)$-plane, with an associated Unruh temperature $T_c=\hbar\Lambda_c/2\pi k_B$, where $\Lambda_c$ is the Lyapunov exponent that characterizes the unstable circular motions of massless particles on the photon sphere. The photon sphere then appears as a location where the thermal bound on chaos for quantum systems with a large number of degrees of freedom, in the form conjectured a few years ago by Maldacena et al., is saturated. The study developed in this paper could hopefully shed a new light on the gravity/CFT correspondence, particularly in asymptotically flat spacetimes, in which the photon sphere 
may also be considered as a holographic screen.
\end{abstract}

\maketitle


\section{\label{sec:Intro} Introduction}
The intriguing duality between black hole (BH) physics and a conformal field theory (CFT) at finite temperature has been investigated for decades, first in the context of the (A)dS/CFT correspondence, e.g. \cite{Maldacena1998, Strominger1999, Strominger2001}, then followed by the Kerr/CFT correspondence \cite{CastroMaloneyStrominger2010} and its extensions (see e.g. \cite{Compere2012} for a complete review and references therein). In most of the Kerr/CFT correspondence extensions \cite{BirminghamGuptaSen2001, GuptaSen2002, BertiniCacciatoriKlemm2012, LoweSkanata2012, LoweMessamahSkanata2014}, among which those better adapted to the case of non-rotating BH spacetimes, the expected conformal structure needed to describe the BH physics through a dual CFT is not found in the symmetry of spacetime itself, but rather as a dynamical symmetry hidden in the equations of motion of the field used to probe the geometry of spacetime. 
More precisely, given a BH geometry, one of the key elements is mainly to explore the behavior of the equations of motion of the considered field theory under the \emph{near horizon limit}, which seems to play a central role in the gravity/CFT duality \cite{Solodukhin1999, Carlip1999, MorettiPinamonti2003}. As an example, it has been shown for the Schwarzschild BH that the dynamics of the field in the \emph{near-horizon limit} tends to be described by an $SL(2,\mathbb{R})$ conformal algebra. Extending the algebra to a Virasoro algebra results in the description of the (near-horizon) BH physics through a dual CFT at finite Hawking temperature defined on the BH horizon. In such a scheme, it is often said that the horizon is playing the role of a holographic screen. However, in the \emph{near-horizon limit}, the use of the hidden dynamical conformal algebra to generate the descendant fields, interpreted as quasinormal modes (QNM), gives only a very specific part of the spectrum of the quasinormal frequencies (QNF): the large overtone behavior of the \emph{highly damped} QNF. Even though this result could be considered as ``incomplete'' (when not questioned \cite{KimMyungPark2013}), it seems to suggest once more that the \emph{highly damped} QNM are related to some physical processes in the near-horizon region of a BH \cite{Maggiore2008, JaramilloMacedoSheikh2021}.\\
 
Surprisingly enough, none of the previous works has been investigating algebraically the origin of an other important part of the QNF spectrum: the \emph{weakly damped} QNF (and their related QNM). In a spirit close to what has been done in the previous cited papers, we suggest a first approach to fill this gap by looking at the physics close to the BH photon sphere (when it exists), and more precisely by introducing and exploring a \emph{near-photon sphere limit} in the study of both a massless scalar field theory and a free-falling test particle in a static and spherically symmetric BH geometry. The main motivation is that the photon sphere, defined as a conformally invariant timelike surface on which a massless particle can orbit the BH on unstable circular null geodesics, plays a prominent role in the study of various phenomena (description of strong gravitational lensing, see e.g \cite{Perlick2004,StefanovYazadjievGyulchev2010,WeiLiuGuo2011,WeiLiu2014,Raffaelli2016} and references therein, reversal of all dynamical effects of rotation \cite{AbramowiczPrasanna1990}), and in particular for the purpose of this paper, in the physical description of the \emph{weakly damped} QNM \cite{CardosoMirandaBertiWitekZanchin2009, DecaniniFolacciRaffaelli2010}. In order to emphasize the existence of a hidden conformal symmetry and the possibility of defining a CFT on the BH photon sphere, this paper will be constructed into two main parts. In the first part, we will consider a massless scalar field theory in a static and spherically symmetric BH spacetime, endowed with a photon sphere. In the \emph{near-photon sphere limit}, we will show that the equation of motion of the scalar field takes a very simple form, with a hidden $SO(2,1) \sim SL(2,\mathbb{R})$ structure that allows to recover algebraically the correct expressions of the weakly damped QNF in the eikonal approximation, and then to construct the associated QNM from the algebra representations. In the second part, we will show that the BH photon sphere acts as an effective Rindler horizon in the $(t,r)$-plane for an ultrarelativistic test particle, coming from infinity, in a non-radial geodesic motion. It will follow that one can associate an Unruh temperature that is related to the Lyapunov exponent characterizing the unstable circular null geodesics on the photon sphere. Remarkably, this temperature corresponds precisely to the saturated thermal bound on chaos for quantum systems with a large number of degrees of freedom, conjectured in \cite{MaldacenaShenkerStanford2016}. From these results, we suggest that the photon sphere, at least for static and spherically symmetric BH, is probably richer than expected: it appears as a particular location where the concepts of instability, thermality and QNM (key ingredients for a dual CFT description) seem to merge naturally, opening hopefully a new path of investigation in the gravity/CFT correspondence.\\

The paper is organized as follows. After the introduction of some notations, we make in Sec.~\ref{sec:Notations} a brief review of a massless particle dynamics, and the associated massless scalar field equation, in a static and spherically symmetric spacetime. We explore in Sec.~\ref{sec:PhotonsphereLimit} the near-photon sphere limit of the massless scalar field equation which enables us to highlight an underlying $SO(2,1) \sim SL(2,\mathbb{R})$ algebra. We then use this algebra in Sec.~\ref{sec:CasimirEigenstates} to recover the weakly damped QNF in the eikonal approximation, and we reconstruct in Sec.~\ref{sec:QNM} the associated QNM from the algebra representations. In Sec.~\ref{sec:ThermalAspects}, from the study of the non-radial geodesic motion in the $(t,r)$-plane of an ultrarelativistic test particle, we show that there should exist thermal effects characterized by a temperature which is related to the Lyapunov exponent that describes the instability of the circular motions of massless particles in the vicinity of the photon sphere. Finally, the paper ends in Sec.~\ref{sec:Cl} with some opening remarks and conclusion.

\section{\label{sec:Notations} Generalities and notations}
\subsection{Black hole metric and massless particle dynamics}\label{subsec:dynamics}
In the Schwarzschild coordinates $(t,r,\theta,\phi)$, we consider a static spherically symmetric four-dimensional spacetime with metric
\begin{equation} \label{metric_BH}
ds^2=-f(r)dt^2+\frac{dr^2}{f(r)}+r^2d\sigma^{2}.
\end{equation}
Here $d\sigma^{2}=d\theta^2 + \sin^2 \theta d\varphi^2$ denotes the line element on the unit 2-sphere $S^{2}$, with $\theta \in [0,\pi]$ and $\varphi \in [0,2\pi]$. Let us consider $r \in ]r_h,+\infty[$, where $r=r_h$ is a simple root of $f(r)$ and defines the location of the BH event horizon. Moreover, let us assume that $f(r)>0$ for $r>r_h$ and $\underset{r \to +\infty}{\lim} f(r) = 1 $. In other words, the considered background geometry is asymptotically flat and the tortoise coordinate $r_\ast = r_\ast(r)$, defined by the relation $dr_\ast/dr = 1/f(r)$, provides a bijection from $]r_h,+\infty[$ to $] -\infty, +\infty[$.\\

Moreover, because of the spherical symmetry of space, let us consider motions on the equatorial plane $\theta=\pi/2$. A free-falling massless particle moves along null geodesics according to
\begin{equation}\label{NullGeodesics}
-f(r)\left(\frac{dt}{d\lambda}\right)^2+\frac{1}{f(r)}\left(\frac{dr}{d\lambda}\right)^2+r^2\left(\frac{d\varphi}{d\lambda}\right)^2=0,
\end{equation}
where $\lambda$ is an affine parameter. From spacetime symmetries, one can define integrals of motion, i.e. energy $E$ and angular momentum $L$ of the massless particle, associated respectively with the Killing vectors $\partial/\partial t$ and $\partial/\partial \varphi$:
\begin{equation}\label{EL}
f(r)\left(\frac{dt}{d\lambda}\right)=E\quad ; \quad r^2\left(\frac{d\varphi}{d\lambda}\right)=L.
\end{equation}
This allows to deduce the equation of motion from (\ref{NullGeodesics})
\begin{equation}\label{EqMotion}
\left(\frac{dr}{d\lambda}\right)^2+V_{\textrm{eff}}(r)=E^2,
\end{equation}
where the effective potential $V_{\textrm{eff}}$ is defined as
\begin{equation}\label{EffPotential}
V_{\textrm{eff}}(r)=\frac{L^2}{r^2}f(r).
\end{equation}
A photon sphere located at $r=r_c$ corresponds here to a local maximum of $V_{\textrm{eff}}$ at $r=r_c$, which reads
\begin{subequations}
\begin{eqnarray}
\left.\frac{d}{dr}V_{\textrm{eff}}(r)\right|_{r_c}=0 &\Leftrightarrow& \frac{2}{r_c}f_c=f'_c, \label{rc}\\
\left.\frac{d^2}{dr^2}V_{\textrm{eff}}(r)\right|_{r_c}<0 &\Leftrightarrow& f''_c - \frac{2}{r_c^2}f_c<0,
\end{eqnarray}
\end{subequations}
where the subscript ``$c$'' means, above and in the following, that the quantity considered is evaluated at $r=r_c$, and the superscripts `` $'$ '' and `` $''$ '' mean respectively the first and second derivatives with respect to $r$.\\

As a consequence, a massless particle reaches the photon sphere if the turning point of its motion satisfies $E^2 = V_\textrm{eff,c} $, i.e. $L/E = r_c/\sqrt{f_c}$, where $r_c/\sqrt{f_c}=b_c$ is the critical impact parameter for massless particles to reach tangentially the photon sphere, before circling the BH at $r=r_c$. Moreover, at $r=r_c$ one also has
\begin{equation}\label{VeffSecond}
V''_{\textrm{eff},c}=-2\eta_c^2 \frac{L^2}{r_c^4},
\end{equation}
where
\begin{equation}
\eta_c = \frac{1}{2}\sqrt{4f_c-2r_c^2f''_c}.
\end{equation}
The study of instability associated with the circular orbits of massless particles on the photon sphere follows. One can for example write down a second order Taylor series expansion of the effective potential near the photon sphere
\begin{equation}\label{Taylor_EffPotential}
V_{\textrm{eff}}(r) \simeq V_{\textrm{eff},c} + \frac{1}{2}V''_{\textrm{eff},c} (r-r_c)^2,
\end{equation}
such as (\ref{EqMotion}) becomes
\begin{equation}
\left(\frac{dr}{d\lambda}\right)^2+V_{\textrm{eff},c} +\frac{1}{2}V''_{\textrm{eff},c} (r-r_c)^2 = E^2.
\end{equation}
If the turning point is very close to $r=r_c$, one can consider that $E^2 \approx V_{\textrm{eff},c}$. The equation of motion then simplifies to
\begin{equation}
\left(\frac{dr}{d\lambda}\right)^2 +\frac{1}{2}V''_{\textrm{eff},c} (r-r_c)^2 = 0,
\end{equation}
which can be written in another, more explicit, form
\begin{equation}
\left(\frac{dr}{dt}\right)^2 + \frac{V''_{\textrm{eff},c}}{2 \dot{t}^2} (r-r_c)^2 = 0,
\end{equation}
with $\dot{t}=dt/d\lambda \approx E/f_c$ close to $r_c$. Finally, the equation of motion in the vicinity of the photon sphere reads
\begin{equation}
\left(\frac{dr}{dt}\right)^2 - \Lambda_c^2 (r-r_c)^2 = 0,
\end{equation}
where 
\begin{equation}\label{Lyapunov}
|\Lambda_c|=\eta_c \frac{\sqrt{f_c}}{r_c}
\end{equation}
is the Lyapunov exponent associated with the unstable circular motion of a free-falling massless particle near the BH photon sphere. In other words, any deviation $\delta r$ on the trajectory of a massless particle will be described by an exponential growth (with respect to $r_c$) within a ``characteristic time'' $|\Lambda_c|^{-1}$, either towards the BH, or to spatial infinity.

\subsection{Massless scalar field in the BH metric}
The Klein-Gordon equation for a massless scalar field $\Phi$ on a general gravitational background is written as
\begin{equation}
\Box \Phi = g^{\mu\nu} \nabla_\mu \nabla_\nu \Phi = \frac{1}{\sqrt{-g}} \partial_\mu \left(\sqrt{-g} g^{\mu\nu}\partial_\nu \Phi\right)=0.
\end{equation}
If the metric is given by eq.~(\ref{metric_BH}) then, after separation of variables, assuming a harmonic time dependence ($e^{-i\omega t}$) for $\Phi$ and the introduction of the radial partial wave functions $\Phi_{\ell\omega}(r)$ with $\ell=0,1,2,\dots$, the Klein-Gordon equation reduces to the well-known Regge-Wheeler equation

\begin{equation}\label{ReggeWheeler}
\frac{d^2 \Phi_{\ell\omega}}{d r_\ast^2} + Q_{\ell\omega}(r_\ast) \Phi_{\ell\omega} = 0.
\end{equation}
where $r_\ast = r_\ast(r)$ is the tortoise coordinate, introduced previously.\\

\noindent In (\ref{ReggeWheeler}), we have introduced $Q_{\ell\omega}(r_\ast(r)) = \omega^2 - V_{\ell}(r)$, with the Regge-Wheeler potential $V_\ell(r)$ defined as
\begin{equation}
V_{\ell}(r)=f(r)\left[\frac{\ell(\ell+1)}{r^2}+\frac{1}{r}f'(r)\right].
\end{equation}
It should be noted that
\begin{itemize}
\item[-] $\lim_{r \to r_h} V_{\ell}(r)=0$ and $\lim_{r \to +\infty} V_{\ell}(r)=0$.
\item[-] For every $\ell \in \mathbb{N}$, $V_{\ell}(r)$ admits a local maximum at $r=r_{0}(\ell)$ which is close to the photon sphere located at $r=r_c$.
\item[-] In the limit $\ell \gg 1$, $r_0(\ell) \approx r_c$ and $V_{\ell}(r) \approx V_{\textrm{eff}}(r)$.
\end{itemize}
Using the tortoise coordinate, we will denote $(r_\ast)_{0,\ell} = r_\ast(r_0(\ell))$ the location of the maximum of $V_{\ell}(r)$. Let us emphasize that, using the tortoise coordinate, the Regge-Wheeler equation (\ref{ReggeWheeler}) looks like a one-dimensional Schr\"{o}dinger equation
\begin{equation}
-\frac{d^2 \Phi_{\ell\omega}}{d r_\ast^2} + \left(-Q_{\ell\omega}(r_\ast)\right) \Phi_{\ell,\omega} = 0,
\end{equation}
with a potential barrier $-Q_{\ell\omega}(r_\ast)$ whose local maximum is located at $(r_\ast)_{0,\ell}$. As noticed by Chandrasekhar \cite{Chandrasekhar1983}, it allows to use all the well-known techniques associated with the study of scattering by a potential barrier, in a very natural way.

\section{\label{sec:PhotonsphereLimit} Hidden conformal symmetry: the near - photon sphere limit}

Following \cite{SchutzWill1985}, let us consider $Q_{\ell\omega}(r_\ast)$ around the location of its local extremum at $(r_\ast)_{0,\ell}$, i.e. around a location as close to the photon sphere as $\ell \gg 1$. A second order Taylor series expansion gives
\begin{equation}\label{QTaylorSeries}
Q_{\ell\omega}(r_\ast) \approx Q_0(\ell,\omega) + \frac{1}{2}Q^{(2)}_0(\ell) \left(r_\ast - (r_\ast)_{0,\ell} \right)^2,
\end{equation}
where we have used the notation
\begin{equation}
Q_0^{(n)}(\ell,\omega)=\left(\frac{d^n Q_{\ell\omega}(r_\ast)}{dr_\ast^n}\right)_{(r_\ast)_{0,\ell}}.
\end{equation}
For $n \geq 1$, $Q_0^{(n)}(\ell,\omega)=Q_0^{(n)}(\ell)$ is independent of $\omega$. We introduce a new dimensionless variable $x$ and a function $h(\ell,\omega)$ defined by
\begin{eqnarray}\label{VariablesPhotonSphereLimit}
&x& = \left[2Q^{(2)}_0(\ell)\right]^{1/4}\left(r_\ast - (r_\ast)_{0,\ell} \right),\nonumber\\ 
&h&(\ell,\omega)=\frac{Q_0(\ell,\omega)}{\sqrt{2Q^{(2)}_0(\ell)}},
\end{eqnarray}
such that (\ref{ReggeWheeler}) can be written as

\begin{equation}\label{EigenvaluePb}
H\tilde{\Phi}_{\ell\omega}(x) = h(\ell,\omega)\tilde{\Phi}_{\ell\omega}(x)
\end{equation}
where $\tilde{\Phi}_{\ell\omega}(x)=\Phi_{\ell\omega}(r_\ast(x))$ and 
\begin{equation}\label{DimensionlessHamitonian}
H=-\frac{d^2}{dx^2} - \frac{1}{4}x^2
\end{equation}
is the (dimensionless) time-independent Hamiltonian governing the massless scalar field dynamics in the near-photon sphere limit. One recognizes that $H$ can be associated with the Hamiltonian of a non-relativistic quantum particle in an inverted parabolic potential. It should be recalled that, in the study of the \emph{near-horizon limit}, the Hamiltonian governing the massless scalar field dynamics reduces to the DeAlfaro-Fubini-Furlan Hamiltonian of conformal quantum mechanics \cite{DeAlfaroFubiniFurlan1976, BirminghamGuptaSen2001, GuptaSen2002}.\\

Let us begin by factorizing $H$. Through the operators $P=-i\frac{d}{dx}$ and $X=\frac{1}{2}x$, such as $[X,P]=\frac{i}{2}\mathbb{I}$, one can for example introduce ``creation and annihilation operators'' analogues
\begin{equation}\label{CreationAnnihilationOperators}
U_{\pm}=\pm P+X=\mp i\frac{d}{dx}+\frac{1}{2}x,\nonumber
\end{equation} 
with $(U_{\pm})^{\dagger} = U_{\pm} \neq U_{\mp}$ (with respect to the usual scalar product) and $[U_{-},U_{+}]=i\mathbb{I}$. The Hamiltonian $H$ can then be written in terms of $U_{\pm}$
\begin{equation}\label{H_Upm}
H=P^2-X^2=-\frac{1}{2}\left(U_{-}U_{+}+U_{+}U_{-}\right).
\end{equation} 
Similarly, one can introduce the dilation operator $D$ and an operator $S$ (related to special conformal transformations generator $K=\frac{1}{4}X^2$)
\begin{eqnarray}\label{DS_Upm}
&D&=\frac{1}{2}\left(PX+XP\right)=\frac{1}{4}\left(U_{+}^2-U_{-}^2\right)=-\frac{i}{2}\left(x\frac{d}{dx} + \frac{1}{2}\right),\nonumber\\
&S&=P^2+X^2=\frac{1}{2}\left(U_{+}^2+U_{-}^2\right)=-\frac{d^2}{dx^2}+\frac{1}{4}x^2.
\end{eqnarray}
Up to numerical factors, it should be noted that the respective roles of $H$ and $D$ are switched whether one is looking at their expressions in terms of $P$ and $X$ operators, or in terms of $U_{\pm}$ operators. The operator $S$ remains unchanged, as a sum of squared operators. It would then become trivial to write the expressions of $H$, $D$ and $S$ as differential operators in ``$u_{\pm}$''-representations, from their expressions in the $x$-representation.\\

To obtain algebraically, from $H$, $D$ and $S$ in the $x$-representation, the spectrum of the near-photon sphere Hamiltonian $H$, which will be related to the spectrum of the \emph{weakly damped} QNF, one can define the following three operators
\begin{subequations}\label{JOperators}
\begin{eqnarray}
J_1&=&D=-\frac{i}{2}\left(x\frac{d}{dx}+\frac{1}{2}\right),\\ 
J_2&=&-\frac{i}{2}S=\frac{i}{2}\left(\frac{d^2}{dx^2}-\frac{1}{4}x^2\right),\\ 
J_3&=&\frac{i}{2}H=-\frac{i}{2}\left(\frac{d^2}{dx^2}+\frac{1}{4}x^2\right).
\end{eqnarray}
\end{subequations}
It is now a simple task to show that they satisfy the commutation relations of an $SO(2,1)$ algebra
\begin{equation}
[J_1,J_2]=-iJ_3; \quad [J_2,J_3]=iJ_1; \quad [J_3,J_1]=iJ_2.
\end{equation}
Moreover, it is possible to introduce ``ladder operators'' in some different ways. For example, one can define
\begin{equation}\label{J+-Operators}
J_{\pm}=\pm iJ_1-J_2=\frac{i}{2}U_{\pm}^2,
\end{equation}
which, together with $J_3$, satisfy an $SL(2,\mathbb{R})$ algebra
\begin{equation}\label{LadderOp_CommutationRelation}
[J_{+},J_{-}]=-2J_3; \quad [J_3,J_{\pm}]=\pm J_{\pm}.
\end{equation}
It should be noted first that the ladder operators $J_{\pm}$ are not self-adjoint conjugate to each other. In our example here, one has instead $\left(J_\pm\right)^\dagger=-J_\pm$, i.e. anti-self-adjoint operators. Finally, let us also note that one could have obviously constructed $J_3$ and $J_{\pm}$, without any references to the operators $D$ and $S$, more directly from $H$ and $U_{\pm}^2$ (as second order differential operators), satisfying commutation relations $(\ref{LadderOp_CommutationRelation})$.

\section{\label{sec:CasimirEigenstates} Casimir operator and eigenstates of $H$}

The quadratic Casimir operator associated with the $SL(2,\mathbb{R})$ algebra is written as:
\begin{equation}\label{CasimirOp}
J^2=J_3^2-J_1^2-J_2^2=J_3^2\pm J_3 \mp J_{\mp}J_{\pm}.
\end{equation}
As usual, one can check that $J^2$ and $J_3$ commutes, so that we can choose a common eigenbasis $\left\{\ket{j,m}\right\}_{(j,m)}$ such as
\begin{subequations}\label{Eigenstates}
\begin{eqnarray}
J^2 \ket{j,m}&=&j(j-1) \ket{j,m}\label{EigenstatesCasimir},\\
J_3 \ket{j,m}&=&m \ket{j,m}\label{EigenstatesJ3}.
\end{eqnarray}
\end{subequations}
Using $[U_{-},U_{+}]=i\mathbb{I}$ with equations (\ref{H_Upm}), (\ref{DS_Upm}) and (\ref{JOperators}), it is easy to show that $J^2=a_0\mathbb{I}$, with the (well-known) value $a_0=-3/16$. So, one deduces that the Bargmann index $j$ can only take two possible real values $j_{1,2}=\frac{1}{2}\left(1 \mp \sqrt{1+4a_0}\right)$, i.e. $j_1=1/4$ and $j_2=3/4$, which implies that the space of states will be covered by two infinite representations of $SL(2,\mathbb{R})$.
Moreover, since $J^2$ has always the same eigenvalue $j_{1,2}(j_{1,2}-1)=a_0$ for every eigenstates, let us note, from now on, $\left\{\ket{a_0,m}\right\}_m$ the eigenbasis common to $J^2$ and $J_3$.\\

From $[J^2,J_{\pm}]=0$ and (\ref{LadderOp_CommutationRelation}), the sets of vectors $\left\{J_{\pm}\ket{a_0,m}\right\}_m$ also define other possible common eigenbases of $J^2$ and $J_3$, associated with the eigenbases $\left\{\ket{a_0,m}\right\}_m$ through 
\begin{equation} 
J_\pm \ket{a_0,m}=c_{\pm}(a_0,m)\ket{a_0,m\pm 1}, 
\end{equation}
where, from basic results on $SO(2,1) \sim SL(2,\mathbb{R})$ algebra, $c_{\pm}(a_0,m)=\sqrt{m(m\pm 1)-a_0}$. Moreover, it should be noted that from (\ref{CasimirOp}), one has
\begin{equation}\label{DualLadder}
\forall m \in \mathbb{C},\, \bra{a_0,m}J_{\mp}J_{\pm}\ket{a_0,m}=m\left(m\pm 1\right)-a_0,
\end{equation}
recalling that here, $\left(J_\pm\right)^\dagger=-J_\pm$. In other words, in our setting, $\bra{a_0,m}J_{\mp}J_{\pm}\ket{a_0,m}$ being a scalar, it implies that the sets of vectors $\left\{\bra{a_0,m}J_\mp\right\}_m$ and $\left\{J_{\pm}\ket{a_0,m}\right\}_m$ belong to two spaces dual to each other, but with $\left(J_\pm \ket{a_0,m}\right)^{\dagger} \neq \bra{a_0,m}J_\mp$.\\

As usual, giving an eigenstate $\ket{a_0,m}$, one can construct with $J_\pm$ a set of corresponding eigenvectors: $\ket{a_0,m\pm 1}$, $\ket{a_0,m \pm 2},\ldots$ . This construction will stop if there exists $m_0^{(\pm)}$ such that $c_{\pm}(a_0,m_0^{(\pm)})=0$. The two possible solutions are $m_0^{(\pm)}= \mp j_{1,2}$, so that  $J_{\pm}\ket{a_0,\mp j_{1,2}}=0$. The states $\ket{a_0,\pm j_{1,2}}$ then define the \emph{ground states} from which one can now construct, from successive applications of $J_\pm$, the eigenstates of $H=-2iJ_3$ and deduce its eigenvalues.\\
From (\ref{EigenstatesJ3}) and (\ref{LadderOp_CommutationRelation}), one can write for every non-negative integer $k$
\begin{subequations}
\begin{eqnarray}
J_3 \ket{a_0,\pm j_{1,2}}&=&\pm j_{1,2}\ket{a_0,\pm j_{1,2}}, \label{J3Eigenstates}\\
J_3 J_\pm^k\ket{a_0,\pm j_{1,2}}&=&\pm \left(j_{1,2} + k\right) J_\pm^k\ket{a_0,\pm j_{1,2}},
\end{eqnarray}
\end{subequations}
which means for the Hamiltonian $H$
\begin{equation}
H J_\pm^k\ket{a_0,\pm j_{1,2}}=\mp i\left(2k + 2j_{1,2}\right) J_\pm^k\ket{a_0,\pm j_{1,2}}.
\end{equation}
Each eigenstate $J_\pm^k\ket{a_0,\pm j_{1,2}}$ of $H$ is related to the next one by a 2-unit step. With an other realization of the generators of the $SL(2,\mathbb{R})$ Lie algebra involving only first order differential operators, instead of second order ones (\ref{J+-Operators}), it would be expected that each eigenstate is related to the next one by a 1-unit step.\\ 

\noindent For each value $j_{1}$ and $j_{2}$, and for every non-negative integer $k$, one has:
\begin{subequations}\label{H_Eigenvalues}
\begin{eqnarray}
H J_\pm^k\ket{a_0,\pm j_{1}}&=&\mp i\left(2k + \frac{1}{2}\right) J_\pm^k\ket{a_0,\pm j_{1}},\\
H J_\pm^k\ket{a_0,\pm j_{2}}&=&\mp i\left(2k + 1 + \frac{1}{2}\right) J_\pm^k\ket{a_0,\pm j_{2}}.
\end{eqnarray}
\end{subequations}
From $(\ref{H_Eigenvalues})$, the eigenstates $J_\pm^k\ket{a_0,\pm j_{1}}$ describe what we shall call the \emph{even} eigenstates, simply noted $\ket{2k,\pm}$, corresponding to the discrete principal series $D^{\pm}(\pm j_{1})$ of the non-unitary irreducible representations of $SO(2,1) \sim SL(2,\mathbb{R})$, and the eigenstates $J_\pm^k\ket{a_0,\pm j_{2}}$ describe the \emph{odd} eigenstates, simply noted $\ket{2k+1,\pm}$, corresponding to the discrete principal series $D^{\pm}(\pm j_{2})$. Therefore, the entire spectrum of $H$ is obtained by considering its eigenspaces as the direct sum of pairs of irreducible representations. Here, there exist two distinct eigenspaces $\mathcal{H}_{+}=D^{+}(+j_{1}) \oplus D^{+}(+j_{2})$ and $\mathcal{H}_{-}=D^{-}(-j_{1}) \oplus D^{-}(-j_{2})$ related to two families of eigenstates, noted $\ket{n,\pm} \in \mathcal{H}_{\pm}$, for every non-negative integer $n$. One can introduce projectors $\Pi_\textrm{even}^{\pm}$ and $\Pi_\textrm{odd}^{\pm}$ on even and odd subspaces of $\mathcal{H}_{\pm}$ respectively, such as $\ket{2k,\pm}=\Pi_\textrm{even}^{\pm} \ket{n,\pm}$ and $\ket{2k+1,\pm}=\Pi_\textrm{odd}^{\pm} \ket{n,\pm}$. For every non-negative integer $n$, the associated spectrum of $H$ can then be written as
\begin{equation}\label{Spectrum_H}
H \ket{n,\pm}=\mp i \left(n+\frac{1}{2}\right)\ket{n,\pm}.
\end{equation}
It should be noted that in the words of a possible CFT defined on the photon sphere, the intermediate states $\ket{a_0,\pm j_{1,2}}$ (or $\ket{0,\pm}$) would be intermediate ``primary states'' with intermediate ``conformal weights'' analogues $\pm j_{1,2}$ (or $h=1/2$), and we would have just constructed above the ``highest weight representations'' of the $SO(2,1) \sim SL(2,\mathbb{R})$ algebra, with intermediate ``descendant states'' $J_\pm^k\ket{a_0,\pm j_{1,2}}$ (or $\ket{n,\pm}$). 
Moreover, having $(J_{\pm})^{\dagger} \neq J_{\mp}$, the corresponding CFT on the photon sphere, if it exists, would probably not be unitary.\\

Finally, one could have obviously stopped the analysis after the $H$ factorization step and decided to solve the problem with the help of a Heisenberg-like algebra, as for the usual harmonic oscillator. More particularly, it should be noted that there exists a simple relation between the two problems. Indeed, introducing the operator $\mathcal{D}_{\pm}=e^{\pm \frac{\pi}{2}D}$ and using the Baker-Campbell-Hausdorff formula with $[D,P]=\frac{i}{2}P$ and $[D,X]=-\frac{i}{2}X$, one can easily show that:
\begin{eqnarray}\label{IHO_HO}
\mathcal{D}_{\pm}\,H \,\mathcal{D}_{\pm}^{-1} &=& \pm i H_{ho}\nonumber \\
\textrm{and}\quad \tilde\Phi_{ho,n}(x)=\mathcal{D}_{\pm} \tilde\Phi_{n}(x)&=&e^{\pm i \frac{\pi}{8}}\tilde\Phi_{n}(e^{\pm i \frac{\pi}{4}}x),
\end{eqnarray}
where $H_{ho}$ is the Hamiltonian of the usual harmonic oscillator and $\tilde\Phi_{ho,n}(x)$ are the associated eigenstates, $\tilde\Phi_{n}(x)$ being the eigenstates of $H$.\\ 

Moreover, although we have noted that $(J_{\pm})^{\dagger}\neq J_{\mp}$ and that $\left\{\bra{a_0,m}J_\mp\right\}_m$ and $\left\{J_{\pm}\ket{a_0,m}\right\}_m$ belong to two spaces dual to each other, we did not discuss on purpose in this paper, the Hilbert space structure of $\mathcal{H}_{\pm}$. Indeed, even though $H$ is self-adjoint (with respect to the usual scalar product), its eigenvalues are found to be purely imaginary. The associated eigenstates $\ket{n,\pm}$ actually belong to the family of Gamow states, lying outside the standard Hilbert space. This has already been extensively treated (see e.g. \cite{CastagninoDienerLaraPuccini1997,Chruscinski2003,Chruscinski2004,CivitareseGadella2004}), and can be analyzed, for the interested reader, in the rigged Hilbert space formulation of quantum physics \cite{Bohm1981}.\\

\section{\label{sec:QNM} From descendant states to quasinormal modes}
\subsection{Quasinormal frequencies}
\noindent Comparing (\ref{Spectrum_H}) and (\ref{EigenvaluePb}), it should exist, in the $x$-representation, two sets of scalar modes $\tilde{\Phi}_{\ell\omega}^{(\pm)}(x)=\braket{x}{n,\pm}$, with $\omega$, $\ell$ and $n$ related by
\begin{equation}\label{WKBEikonal}
h(\ell,\omega)=\mp i\left(n+\frac{1}{2}\right).
\end{equation}
To solve this equation, one has to go into the complex $\omega$-plane, for each integer $\ell$. One could also solve this equation in the complex angular momentum plane, keeping $\omega>0$, as it will be discussed in Sec.~\ref{sec:Cl}. It should be noted that (\ref{WKBEikonal}) corresponds to the analytic WKB connection formula in the eikonal approximation, introduced in \cite{SchutzWill1985}. Then, for every non-negative integer $n$, the solutions $\omega_{\ell n}$ correspond to the family of quasinormal frequencies in the eikonal approximation, for the static and spherically symmetric BH described by (\ref{metric_BH}). Solving $(\ref{WKBEikonal})$ for $\ell \gg 1$ gives the well-known expressions
\begin{equation}\label{QNF}
\omega_{\ell n}^{(\pm)}=\pm\frac{\sqrt{f_c}}{r_c}\ell-i\eta_c \frac{\sqrt{f_c}}{r_c}\left(n+\frac{1}{2}\right).
\end{equation}
The $(\mp)$ sign in (\ref{WKBEikonal}), or $(\pm)$ in (\ref{QNF}), is associated with the eigenspaces $\mathcal{H}_{\pm}$, and is also related to the location of the solutions in the complex $\omega$-plane. More precisely, the solutions $\omega_{\ell n}^{(\pm)}$ are located in the third and fourth quadrants of the complex $\omega$-plane. These known symmetries are described in detail in \cite{Iyer1987}. Moreover, $\textrm{Im}(\omega_{\ell n})<0$ allows to avoid any instability for the field $\Phi$. In the following, let us simply note $\tilde{\Phi}_{\ell\omega}^{(\pm)}(x)=\tilde{\Phi}_{n}^{(\pm)}(x)$ the scalar modes.

\subsection{Quasinormal modes}

The intermediate ``primary states'' $\ket{a_0,\pm j_{1,2}}$ with intermediate ``conformal weights'' $\pm j_{1,2}$ are ground states of the $J_{\mp}$ ladder operators. It should correspond intermediate ``primary fields'' $\varphi_{0}^{(\pm)}(x)$ in the $x$-representation such as $J_{\mp} \varphi_{0}^{(\pm)}(x) = 0$. Let us recall that in this section, the superscripts $(\pm)$ between brackets is associated with the sign of the ``conformal weights'' $\pm j_{1,2}$, i.e. related to $\mathcal{H}_{\pm}$. From (\ref{J+-Operators}), the ``primary fields'' should then satisfy:
\begin{equation}
\left(U_{\mp}\right)^2 \varphi_{0}^{(\pm)}(x) = 0.
\end{equation} 
As a second order differential equation in the $x$-representation, there are two possible solutions $\varphi_{0,1}^{(\pm)}(x)$ and $\varphi_{0,2}^{(\pm)}(x)=U_{\pm}\varphi_{0,1}^{(\pm)}(x) \neq 0$ such that
\begin{subequations}
\begin{eqnarray}
U_{\mp} \varphi_{0,1}^{(\pm)}(x) &=& 0,\label{PrimaryField01} \\
\left(U_{\mp}\right)^2 \varphi_{0,2}^{(\pm)}(x) &=& 0.
\end{eqnarray}
\end{subequations}
Using the commutation relation $[U_{-},U_{+}]=i\mathbb{I}$, one can show that the primary fields also satisfy
\begin{subequations}\label{PrimaryFieldsProperties}
\begin{eqnarray}
U_{\mp} \varphi_{0,2}^{(\pm)}(x) &=& \pm i \varphi_{0,1}^{(\pm)}(x),\\
U_{\mp}U_{\pm} \varphi_{0,1}^{(\pm)}(x) &=& \pm i \varphi_{0,1}^{(\pm)}(x),\\
U_{\pm}U_{\mp} \varphi_{0,2}^{(\pm)}(x) &=& \pm i \varphi_{0,2}^{(\pm)}(x).
\end{eqnarray}
\end{subequations}

\noindent Finally, from (\ref{PrimaryFieldsProperties}), (\ref{J3Eigenstates}) and writing $J_3=(-i/2)U_{\mp}U_{\pm}\mp 1/4$, one can check that $\varphi_{0,1}^{(\pm)}$ corresponds to $\ket{a_0,\pm j_{1}}$, and $\varphi_{0,2}^{(\pm)}$ to $\ket{a_0,\pm j_{2}}$, with ``conformal weights'' $\pm j_1$ and $\pm j_2$ respectively.\\

Solving (\ref{PrimaryField01}), one has
\begin{subequations}
\begin{eqnarray}
\varphi_{0,1}^{(\pm)}(x) &=& a_{\pm}e^{\pm ix^2/4},\\
\varphi_{0,2}^{(\pm)}(x) &=& x \varphi_{0,1}^{(\pm)}(x),
\end{eqnarray}
\end{subequations}
with $a_{\pm}$ a constant. Up to a normalization constant, the descendant states $\varphi_{k,1}^{(\pm)}(x)\propto J_{\pm}^k \varphi_{0,1}^{(\pm)}(x)$ and $\varphi_{k,2}^{(\pm)}(x) \propto J_{\pm}^k \varphi_{0,2}^{(\pm)}(x)$ can be computed by successive iterations, and one obtains for every non-negative integer $k$

\begin{subequations}
\begin{eqnarray}
\varphi_{k,1}^{(+)}(x) &=& a_{k}^{(+)} H_{2k}(\mp e^{-i\pi/4}x)e^{+ ix^2/4},\\
\varphi_{k,1}^{(-)}(x) &=& a_{k}^{(-)} H_{2k}(\mp e^{+i\pi/4}x)e^{- ix^2/4},\\
\varphi_{k,2}^{(+)}(x) &=& b_{k}^{(+)} H_{2k+1}(\mp e^{- i\pi/4}x)e^{+ ix^2/4},\\
\varphi_{k,2}^{(-)}(x) &=& b_{k}^{(-)} H_{2k+1}(\mp e^{+ i\pi/4}x)e^{- ix^2/4},
\end{eqnarray}
\end{subequations}
where $H_k(X)$ is a $k^{th}$-order Hermite polynomial of the $X$ variable, and $a_{k}^{(\pm)}$ and $b_{k}^{(\pm)}$ are functions of $k$, which are not relevant here to discuss the behavior of the descendant fields as functions of $x$. It should be noted that, for each case, the Hermite polynomials can be obtained from two possible change of variables $X_{\mp}=\mp e^{- i\pi/4}x$ for $\varphi^{(+)}_{k}$, and $X_{\mp}=\mp e^{+ i\pi/4}x$ for $\varphi^{(-)}_{k}$. One can then deduce the descendant fields $\tilde{\Phi}_{n}^{(\pm)}(x)$ for every non-negative integer $n$ as a direct sum on the even and odd subspaces of $\mathcal{H}_{\pm}$, with the help of the corresponding projectors. By noting $\tilde{\Phi}^{(+)}_{n}(x)=\tilde{\Phi}^{\textrm{out}}_{n,\pm}(x)$ and $\tilde{\Phi}^{(-)}_{n}(x)=\tilde{\Phi}^{\textrm{in}}_{n,\pm}(x)$, where the subscripts ``$\pm$'' are related to $X_{\mp}$, one obtains, up to normalization constants $f_n^{\textrm{out}}$ and $g_n^{\textrm{in}}$:
\begin{subequations}
\begin{eqnarray}
\tilde{\Phi}^{\textrm{out}}_{n,\pm}(x) &=& f_n^{\textrm{out}} H_n(\mp e^{i\pi/4}x) e^{+ ix^2/4},\\
\tilde{\Phi}^{\textrm{in}}_{n,\pm}(x) &=& g_n^{\textrm{in}} H_n(\mp e^{-i\pi/4}x) e^{- ix^2/4}.
\end{eqnarray}
\end{subequations}
The descendant states $\tilde{\Phi}^{\textrm{out}}_{n,\pm}$ are related to $\mathcal{H}_{+}$, i.e. associated with eigenvalues $h(\ell,\omega)=-i(n+1/2)$ of $H$, and $\tilde{\Phi}^{\textrm{in}}_{n,\pm}$ are related to $\mathcal{H}_{-}$, i.e. associated with eigenvalues $h(\ell,\omega)=+i(n+1/2)$ of $H$. Re-introducing the time dependence $e^{-i\omega t}$, with $\omega$ a solution of (\ref{WKBEikonal}), one can interpret, for each $n$, the modes $\tilde{\Phi}^{\textrm{out}}_{n,\pm}$ as two purely outgoing waves (from the peak of the Regge-Wheeler potential), exponentially decaying with time as long as they move away from the peak of the potential. The modes $\tilde{\Phi}^{\textrm{in}}_{n,\pm}$ can be interpreted as two purely ingoing waves (towards the peak of the Regge-Wheeler potential), exponentially decaying in time as long as they get close to the peak of the potential. From a physical point of view, $\tilde{\Phi}^{\textrm{out}}_{n,\pm}$ actually are the physical states and define precisely the quasinormal modes of the BH.

\section{\label{sec:ThermalAspects} Thermal aspects in the vicinity of the photon sphere}
\subsection{Near-photon sphere limit, effective geodesic motion and Rindler metric approximation}

In this section, we probe the vicinity of the BH photon sphere from a kinematical point of view by considering the motion of a free-falling test particle of mass $m$ following the geodesic line element (\ref{metric_BH}). As it will be shown below, for a massive test particle to get very close to the photon sphere, we will need to consider the ultrarelativistic limit of its motion.\\

From the spherical symmetry, and without loss of generality, let us write the line element (\ref{metric_BH}) associated with the timelike motion of the test particle in the equatorial plane $\theta=\pi/2$ (equatorial planes extremize the test particle action in the gravitational field, with respect to $\theta$)
\begin{equation}\label{geodesics_equatorial_plane}
ds^2=-f(r)dt^2+\frac{dr^2}{f(r)}+r^2d\varphi^{2}.
\end{equation}
In the massive case, one can define the integrals of motion, related to the Killing vectors $\partial/\partial t$ and $\partial/\partial \varphi$, which read here 
\begin{equation}\label{EL_massive}
f(r)\left(\frac{dt}{d\tau}\right)=\frac{E}{m}\quad ; \quad r^2\left(\frac{d\varphi}{d\tau}\right)=\frac{L}{m}
\end{equation}
where $\tau$ is the particle proper time. The energy $E$ being a constant of motion, one can relate $E$ at spatial infinity with the test particle momentum $p$ and mass $m$
\begin{equation}
p(E)=\sqrt{E^2-m^2}.
\end{equation}
At spatial infinity, the test particle velocity is
\begin{equation}
v(E)=\frac{p(E)}{E}=\sqrt{1-\frac{m^2}{E^2}}.
\end{equation}
The angular momentum $L$ being also a constant of motion, one can write at spatial infinity
\begin{equation}
L=p(E)b
\end{equation}
where $b$ is the impact parameter of the test particle, free-falling towards the BH, which allows us to write the impact parameter $b$ of the test particle as
\begin{equation}
b=\frac{L}{Ev(E)}.
\end{equation}
It should be noted that for a massless particle ($m=0$ and $v(E)=1$), one simply has $p(E)=E$ and $b_0=L/E < b$. As shown in Sec.~\ref{subsec:dynamics}, the critical impact parameter for massless particles to reach tangentially the photon sphere, is obtained for values of $L$ and $E$ such that $E^2=V_{\textrm{eff},c}$, which reads $b_c=L/E=r_c/\sqrt{f_c}$.\\

Using (\ref{EL_massive}), one can deduce the equation of motion for the test particle
\begin{equation}\label{EqMotionMassive}
m^2\left(\frac{dr}{d\tau}\right)^2 + U_\textrm{eff}(r) = E^2,
\end{equation}
where 
\begin{equation}
U_\textrm{eff}(r)=f(r)\left[\frac{L^2}{r^2}+m^2\right].
\end{equation}
The locations $r=r_i$ of the extrema of $U_\textrm{eff}(r)$ satisfy
\begin{equation}\label{r0}
f'(r_i)-\frac{2}{r_i^2}f(r_i)+\frac{m^2}{L^2}f'(r_i)=0.
\end{equation}
Let us assume that the possible values of the particle angular momentum $L$ are such that $U_\textrm{eff}(r)$ admits a local maximum. Let us call $r_0(L)$ the location of this local maximum. In the limit $L \gg 1$, eq.~(\ref{r0}) tends to eq.~(\ref{rc}), the local maximum coincides with the location of the photon sphere, i.e. $r_0(L)$ tends to $r_c$ and $U_\textrm{eff}(r_0(L))$ tends to $V_\textrm{eff,c}$.\\

Therefore, a test particle (with energy $E$ and angular momentum $L$) coming from infinity, gets very close to the photon sphere if at least
\begin{subequations}\label{EL_conditions}
\begin{eqnarray}
&m^2& < E^2 \approx U_\textrm{eff}(r_0(L)),\label{E_condition}\\
&L& \gg 1 \quad \textrm{with} \quad L/E \quad \textrm{finite} \label{L_condition}.
\end{eqnarray}
\end{subequations}
The condition (\ref{E_condition}) implies that the turning point of the particle motion is located in the vicinity of the maximum of $U_\textrm{eff}(r)$. The second condition (\ref{L_condition}) implies that the location of this maximum tends to the location of the photon sphere. If the conditions (\ref{EL_conditions}) are both satisfied, then one has $E^2 \approx U_\textrm{eff}(r_0(L)) \approx V_\textrm{eff,c}$, i.e. $L/E \approx r_c/\sqrt{f_c}$, which corresponds to the limit $v(E) \to 1$ of an ultrarelativistic test particle. In other words, the impact parameter $b$ tends, from above, to the critical impact parameter $b_c$ associated with massless particles motion, and the particle coming from infinity will get close to the photon sphere before moving away, back to infinity.\\
 
Let us now focus on the near-photon sphere limit of (\ref{geodesics_equatorial_plane}) to describe the motion of the test particle in the ultrarelativistic limit. Let us first use the symmetry conditions (\ref{EL_massive}) for geodesic motions \cite{Rindler2006}, to restrict ourselves to an equivalent effective geodesic line element in the $(t,r)$-plane. From (\ref{EL_massive}), one can write
\begin{equation}
r^2 d\varphi^2=f(r)^2\frac{L^2}{E^2 r^2}dt^2.
\end{equation} 
This allows to transform (\ref{geodesics_equatorial_plane}) into the following line element, that would obviously give the same radial equation of motion (\ref{EqMotionMassive}),
\begin{equation}\label{metric_BH2}
ds^2=f(r)\left(-1+\frac{V_\textrm{eff}(r)}{E^2}\right)dt^2+\frac{dr^2}{f(r)},
\end{equation}
where $V_\textrm{eff}(r)$ is still formally defined by expression (\ref{EffPotential}) but with $L$ being now the angular momentum of the test particle. It should be noted that the line elements (\ref{metric_BH2}) and (\ref{geodesics_equatorial_plane}) have the same magnitude for any geodesic motion, i.e. for any given $E$ and $L$. Moreover, let us emphasize that (\ref{metric_BH2}) does not describe a purely radial motion in the BH background (in such case $\varphi$ would have been constant, i.e. $L=0$, and the photon sphere would have had no effect on the test particle motion), but rather describes the effective non-radial geodesic motion of a test particle in the $(t,r)$-plane of a static and spherically symmetric BH background, taking into account explicitly the effect of the centrifugal potential barrier in its time component. Let us recall that we will not consider the case where the particle gets trapped into the BH, i.e. here one has $b \gtrsim b_c$.\\ 

The near-photon sphere limit of (\ref{metric_BH2}) requires the conditions (\ref{EL_conditions}) to be satisfied, i.e. $E^2 \approx V_\textrm{eff,c}$, and is obtained from the lowest order Taylor series expansion around $r=r_c$ which does not cancel the time component of (\ref{metric_BH2}). Using (\ref{Taylor_EffPotential}), the effective line element (\ref{metric_BH2}) then becomes in this limit
\begin{equation}
ds^2 \simeq -\frac{V''_\textrm{eff,c}}{2E^2}f_c(r-r_c)^2dt^2+\frac{dr^2}{f_c}.
\end{equation}
From (\ref{VeffSecond}) and (\ref{Lyapunov}), and introducing the variable
\begin{equation}
\rho = \frac{r-r_c}{\sqrt{f_c}} \Leftrightarrow d\rho = \frac{dr}{\sqrt{f_c}},
\end{equation}
one obtains a simple Rindler form of the effective line element near the photon sphere, which acts in this setting as an effective Rindler horizon
\begin{equation}\label{Rindler_metric}
ds^2 \simeq - \Lambda_c^2 \rho^2 dt^2 + d\rho^2,
\end{equation}
where we have used $L^2/E^2 \approx r_c^2/f_c$ because $E^2 \approx V_\textrm{eff,c}$.\\
The Lyapunov exponent $|\Lambda_c|$ associated with the massless particle motions around the photon sphere plays the role of a constant proper acceleration in the near-photon sphere limit of the line element (\ref{metric_BH2}), describing the test particle effective geodesic motion in the $(t,r)$-plane, a role analogue to the role played by the surface gravity in the near-horizon limit. The thermal aspects then follow directly. 

\subsection{\label{sec:Thermalization} Scalar field thermalization in the vicinity of the photon sphere}
From the Rindler metric (\ref{Rindler_metric}), the link with thermal aspects of a scalar field theory in the vicinity of the photon sphere can then be obtained as usual, in different ways, i.e. through Bogolyubov transformations between vacuum states or Euclidean quantum field theory approach. Inspired by \cite{AlsingMilonni2004}, we choose to follow a route closely related to the first approach.\\

In the following, let us note $|\Lambda_c|=\Lambda_c$ for short. Let us then consider the change of variables
\begin{subequations}
\begin{eqnarray}
T&=&\rho \sinh(\Lambda_c t),\\
R&=&\rho \cosh(\Lambda_c t)
\end{eqnarray}
\end{subequations}
such that the effective line element (\ref{Rindler_metric2}) takes a local Minkowski form near the photon sphere
\begin{equation}\label{Rindler_metric21}
ds^2 = - dT^2 + dR^2.
\end{equation}

\noindent Let us now define the variable $\xi$ such as
\begin{equation}
\rho = \frac{1}{\Lambda_c} e^{\Lambda_c \xi}.
\end{equation} 
It should be noted that, with this change of variable, the location of the photon sphere is being pushed away to $\xi \to -\infty$. The metric (\ref{Rindler_metric}) then reads
\begin{equation}\label{Rindler_metric2}
ds^2=e^{2\Lambda_c \xi}\left[-dt^2+d\xi^2\right].
\end{equation}
The last step is to introduce two couples of variables, the first one, $(u,v)$, such as 
\begin{subequations}\label{uv}
\begin{eqnarray}
u&=&t+\xi\\
v&=&t-\xi
\end{eqnarray}
\end{subequations}
in which the metric becomes
\begin{equation}\label{Rindler_metric3}
ds^2=-e^{2\Lambda_c \xi}dudv,
\end{equation}
and the second one, $(U,V)$, such as
\begin{subequations}\label{UV}
\begin{eqnarray}
U&=&T-R=-\frac{1}{\Lambda_c}e^{-\Lambda_c u}\\
V&=&T+R=\frac{1}{\Lambda_c}e^{\Lambda_c v}.
\end{eqnarray}
\end{subequations}
in which metric $(\ref{Rindler_metric21})$ simply reads
\begin{equation}
ds^2=-dUdV.
\end{equation}
\noindent Let us now consider a massless scalar field $\phi$ in the vicinity of the photon sphere. More precisely, in $(T,R)$ or equivalently in $(U,V)$ coordinates, let us look at a Fourier component (i.e. a local ``plane wave'') of $\phi$, propagating towards (resp. away from) the BH, noted $\phi^{\textrm{in}}$ (resp. $\phi^{\textrm{out}}$) 
\begin{subequations}
\begin{eqnarray}
\phi^{\textrm{out}}_{\Omega,\pm} (U)=\exp\left(\mp i\Omega U\right),\\
\phi^{\textrm{in}}_{\Omega,\pm} (V)=\exp\left(\mp i\Omega V\right).
\end{eqnarray}
\end{subequations}
In the $(u,v)$ coordinate system, it reads
\begin{subequations}
\begin{eqnarray}
\phi^{\textrm{out}}_{\Omega,\pm} (u)&=&\exp\left[\pm i\left(\frac{\Omega}{\Lambda_c}\right) e^{-\Lambda_c u}\right],\\
\phi^{\textrm{in}}_{\Omega,\pm} (v)&=&\exp\left[\mp i\left(\frac{\Omega}{\Lambda_c}\right) e^{\Lambda_c v}\right].
\end{eqnarray}
\end{subequations}

\noindent Focusing on the outgoing component, the Fourier transform of the scalar field seen from an accelerated observer, i.e. the test particle, reads:
\begin{eqnarray}\label{FourierPlaneWave}
\int_{-\infty}^{+\infty} du e^{i\omega u} \phi^{\textrm{out}}_{\Omega,\pm}(u)&=&\frac{1}{\Lambda_c}\int_{0}^{+\infty}dw w^{-i\frac{\omega}{\Lambda_c}-1}e^{\pm i\frac{\Omega}{\Lambda_c}w}\nonumber\\
&=&\frac{1}{\Lambda_c}\left(\frac{\Omega}{\Lambda_c}\right)^{\frac{i\omega}{\Lambda_c}}e^{\pm \frac{\pi \omega}{2\Lambda_c}}\Gamma\left(-i\frac{\omega}{\Lambda_c}\right).\nonumber\\
\end{eqnarray}
In the first line, we have introduced $w=e^{-\Lambda_c u}$. For the resulting integral to converge, it has been regularized considering $\omega \to \omega +i\Lambda_c \varepsilon$ with $0<\varepsilon<1$, taking then the limit $\varepsilon \to 0$. Finally, we used the definition of the gamma function \cite{AbramowitzStegun1965} and $(\mp i)^{i\frac{\omega}{\Lambda_c}}=e^{\pm \frac{\pi \omega}{2\Lambda_c}}$, which is the mathematical key to the Unruh effect.\\

\noindent In the context of Rindler/Minkowski mapping, the Bogolyubov coefficients $\alpha_{\omega \Omega}$ and $\beta_{\omega \Omega}$ are known to be simply the Fourier transforms of $\phi^{\textrm{out}}_{\Omega,+} (u)$ and $\phi^{\textrm{out}}_{\Omega,-} (u)$ respectively
\begin{subequations}
\begin{eqnarray}
\alpha_{\omega \Omega}&=&\frac{1}{2\pi}\int_{-\infty}^{+\infty} du e^{i\omega u} \phi^{\textrm{out}}_{\Omega,+}(u),\\
\beta_{\omega \Omega}&=&-\frac{1}{2\pi}\int_{-\infty}^{+\infty} du e^{i\omega u} \phi^{\textrm{out}}_{\Omega,-}(u).
\end{eqnarray}
\end{subequations}
Then, from (\ref{FourierPlaneWave}), one deduces immediately
\begin{equation}
|\alpha_{\omega \Omega}|^2=e^{\frac{2\pi\omega}{\Lambda_c}}|\beta_{\omega \Omega}|^2.
\end{equation}
With the usual Minkowski and Rindler vacuum states, their related Fock spaces and the normalization condition on Bogolyubov coefficients, the number of quanta seen from the accelerated observer then follows the well-known Bose-Einstein distribution to which corresponds a temperature 
\begin{equation}
T_c=\frac{\Lambda_c}{2\pi}=\eta_c \frac{\sqrt{f_c}}{2\pi r_c}.
\end{equation}
Without any surprise, it should be noted that $T_c < T_H$, where $T_H$ is the Hawking temperature, especially for the Schwarzschild BH for which $T_c/T_H=4/3\sqrt{3}$. Moreover, it is quite surprising and worth noting that we have obtained, from the near-photon sphere limit of the effective geodesic line element (\ref{metric_BH2}), an Unruh temperature $T_c$ which exactly saturates the thermal bound on chaos, in the form conjectured by Maldacena, Shenker and Stanford \cite{MaldacenaShenkerStanford2016}. With the algebraic $SL(2,\mathbb{R})$ approach developed above, those are important clues suggesting that one may probably define a CFT on a BH photon sphere.

\section{\label{sec:Cl} Remarks and conclusion}

In this paper, we have introduced a near-photon sphere limit within which the equation of motion of a massless scalar field in a static and spherically symmetric BH spacetime takes a very simple form, and exhibits a $SL(2,\mathbb{R})$ algebraic structure. From the $SL(2,\mathbb{R})$ algebra and its representations, we have computed the correct expressions of the weakly damped QNF in the eikonal approximation and show that the associated states satisfy the BH QNM conditions. This approach encompasses a large class of BH endowed with a photon sphere (Schwarzschild, Reissner-Nordstr\"{o}m, Schwarzschild-de Sitter, canonical acoustic BH) and remains valid as long as the effective potential associated with the BH admits a finite maximum. Moreover, from the constants of motion, we have considered an effective line element describing the non-radial geodesic motion of a test particle in the $(t,r)$-plane, coming from spatial infinity and moving towards the BH with an impact parameter $b \gtrsim b_c$. In the near-photon sphere limit, this line element interestingly reduces to a Rindler line element from which on can compute an Unruh temperature $T_c=\hbar \Lambda_c/2\pi k_B$ for the associated scalar field theory, that exactly saturates the thermal bound on chaos. The approach developed in this paper can of course be generalized in a straightforward way to any number of spacetime dimensions. It would also be of great interest to have a closer look at the Kerr BH case, for which, due to the axial symmetry, there exist two photon circular orbits (prograde and retrograde) in the equatorial plane of the BH. Outside the equatorial plane, the photon spherical orbits structure is more subtle (see e.g. \cite{Teo2003} and references therein).\\ 

With the photon sphere temperature $T_c$, the eikonal approximation of the \emph{weakly damped} QNF takes a more ``universal'' form in line with a dual CFT description:
\begin{equation}\label{QNF_Temperature}
\omega_{\ell n}^{(\pm)}= \pm \Omega_c \ell - 2\pi i T_c \left(n+\frac{1}{2}\right)
\end{equation}
where we have introduced $\Omega_c = 2\pi/\tau_c$ the angular velocity of massless particles on circular orbits, with $\tau_c = 2\pi r_c/\sqrt{f_c}$ the time that a massless particle takes to circle the BH at $r=r_c$. Although being given in the eikonal approximation, the form (\ref{QNF_Temperature}) is highly suggestive if one thinks about QNF as poles, in the momentum representation, of a scalar 2-point correlation function of an underlying CFT. Moreover, by introducing second order differential operators realizations for the $SL(2,\mathbb{R})$ algebra, we have obtained intermediate ``conformal weights'' analogues, namely $j_{1,2}$. We naturally expect that the conformal weights from a complete Virasoro algebra approach, associated to BH with metric (\ref{metric_BH}), will be given by $h=\bar{h}=j_2-j_1=1/2$ ($h \neq \bar{h}$ for spinning BH), as it is suggested by eq.~(\ref{Spectrum_H}) or by using the Legendre duplication formula for gamma functions \cite{AbramowitzStegun1965} on the expression of the thermal CFT scalar 2-point function in momentum space.\\

The photon sphere appears as a particular location, at least for any static and spherically symmetric BH geometries, where the concepts of instability, thermality and quasinormal modes seem to meet naturally. In other words, one can naturally find the key ingredients to define a CFT at finite temperature on the photon sphere. It would be interesting to pursue this investigation further by looking first at the hidden Virasoro algebra involving first order operators, then to focus on the energy-momentum tensor of the CFT, the associated central charge and Cardy entropy. Hopefully, this work could shed a new light on the gravity/CFT correspondence for non-AdS spacetimes. In the AdS/CFT case, the AdS conformal boundary is enough to describe the entire bulk, due to the conformal symmetry of the spacetime geometry. In non-AdS spacetimes, hidden conformal symmetries can be found near the BH horizon as well as near its photon sphere which in turn could be both considered as holographic screens, respectively at Hawking temperature and Unruh temperature $T_c$, and probably describing at least, two different regions in the bulk of the BH geometry. This question remains open.\\

The Unruh temperature $T_c$ put aside, this work also brings into light a new pedagogical way of studying resonant scattering problems in any one dimensional setting from the top of an effective potential barrier from a purely algebraic point of view, and make a bridge between the well-known analytical WKB method and a corresponding algebraic approach that gives a WKB connection-like formula, at least at the eikonal level, without using any matching of solutions. Moreover, the inverted harmonic oscillator, which is described by the Hamiltonian we have found in the near-photon sphere limit, turns out to be a very simple but very rich toy model allowing to capture the essence of various phenomena, from condensed matter to BH physics \cite{SubramanyanHedgeVishveshwaraBradlyn2019}.\\

Finally, it should be noted that eq.~(\ref{WKBEikonal}) could have also been solved in the complex $\lambda$-plane, where here $\lambda=\ell+1/2$. The complex $\lambda$-plane is used in the framework of the complex angular momentum theory, an asymptotic approach to resonant scattering problems. The solution of eq.~(\ref{WKBEikonal}) then gives the expression (here, in the eikonal approximation) of the Regge poles $\lambda_n(\omega)$ (with $\omega>0$) of the associated S-matrix that describes the scattering by a BH (see \cite{DecaniniFolacciRaffaelli2010} for a more detailed description). In a few words, the resonant scattering of a given field theory by a BH, is described by the Regge poles of the corresponding S-matrix (or of the diffractive part of the related Feynman propagator for non asymptotically flat BH geometries). It has been shown that each Regge pole is associated with the propagation of a surface wave on the BH photon sphere. In particular, a resonance, i.e. a quasinormal mode (or, more precisely here, a ``Regge mode''), is understood as a Breit-Wigner resonance produced by constructive interferences between different components of the associated surface wave, each component corresponding to a different number of circumvolutions of the wave around the photon sphere. For the purpose of this paper and in the Regge pole framework, the example of the Ba\~{n}ados-Teitelboim-Zanelli (BTZ) BH is interesting, even though it would appear in our setting as a singular case. On the one hand, from the point of view of the AdS/CFT correspondence, it has been shown \cite{BirminghamSachsSolodhukin2002} that the QNF spectrum of the BTZ BH can be computed as poles in the momentum representation of the retarted 2-point correlation function of an underlying 2-dimensional CFT, defined on its boundary. On the other hand, the Regge poles description of QNM in terms of surface waves remains the same \cite{DecaniniFolacci2009}, the surface waves being supported by the BTZ BH boundary at infinity, which actually acts as a photon sphere. From this example, it would then be naturally expected, as an extension of this paper, that correlation functions of the hidden CFT defined on any BH photon sphere would have a simple Regge poles description, and in particular that the scalar 2-point correlation function at high energy would have the same behavior as the BH high energy absorption cross section (see e.g. \cite{DecaniniEspositoFareseFolacci2011}). From a physical point of view, it would not be surprising that, in the framework of Regge poles, the related surface waves would be describing the relaxing process of a perturbed thermal state going back to equilibrium in a CFT defined on the photon sphere, the imaginary part of the Regge poles being inversely proportional to a characteristic absorption length of the relaxing process. Finally, to complete the study of the CFT correspondence on a BH photon sphere in the Regge poles framework, it might be expected that a Virasoro algebra originates from a Korteveg de Vries-like equation \cite{LenziSopuerta2021} that would hopefully describe the propagation of the corresponding surface waves on the photon sphere. The Regge poles description of the CFT correspondence on a BH photon sphere is left to future work.
  
\begin{acknowledgments}
The author would like to thank J.P. Provost and J.L. Jaramillo for stimulating discussions, and R. Ruffini for constant support. The IMB receives support from the EIPHI Graduate School (contract ANR-17-EURE-0002).
\end{acknowledgments}





\bibliography{BR_HiddenCFTPhotonSphere_bib}

\end{document}